\documentclass[aps,prl,groupedaddress,showpacs,nofootinbib]{revtex4}
\usepackage{graphicx}
\usepackage{times}
\def\beq{\begin{equation}}
\def\eeq{\end{equation}}
\def\bey{\begin{eqnarray}}
\def\eey{\end{eqnarray}}

\def\lsim{\mathrel{\raise.3ex\hbox{$<$\kern-.75em\lower1ex\hbox{$\sim$}}}}
\def\gsim{\mathrel{\raise.3ex\hbox{$>$\kern-.75em\lower1ex\hbox{$\sim$}}}}

\begin{document}

\title{Dark Matter and Gamma-Rays From Draco: MAGIC, GLAST and CACTUS}
\author{Lars Bergstr\"om$^{1}$ and Dan Hooper$^{2}$}
\address{$^1$Department of Physics, Stockholm University, AlbaNova University Center, SE-106 91 Stockholm, Sweden\\$^2$Fermi National Accelerator Laboratory, Particle Astrophysics Center, Batavia, IL  60510}

\date{\today}

\begin{abstract}
The dwarf spheroidal galaxy Draco has long been considered likely to be one of the brightest point sources of gamma-rays generated through dark matter annihilations. Recent studies of this object have found that it remains largely intact from tidal striping, and may be more massive than previously thought.  In this article, we revisit Draco as a source of dark matter annihilation radiation, with these new observational constraints in mind. We discuss the prospects for the experiments MAGIC and GLAST to detect dark matter in Draco, as well as constraints from the observations of EGRET. We also discuss the possibility that the CACTUS experiment has already detected gamma-rays from Draco. We find that it is difficult to generate the flux reported by CACTUS without resorting to non-thermally produced WIMPs and/or a density spike in Draco's dark matter distribution due to the presence of an intermediate mass black hole. We also find that for most annihilation modes, a positive detection of Draco by CACTUS would be inconsistent with the lack of events seen by EGRET.
\end{abstract}
\pacs{95.35.+d;95.30.Cq,98.52.Wz,95.55.Ka
\hspace{0.5cm} FERMILAB-PUB-05-538-A}
\maketitle

\section{Introduction}

It has long been thought that dark matter particles could be observed indirectly by detecting the products of their annihilations. Such products, including gamma-rays, neutrinos and anti-matter, have been searched for using a wide range of experimental techniques~\cite{review}. Gamma-rays from dark matter annihilations, in particular, have been sought after using both satellite and ground based experiments.

The potential astrophysical source of gamma-rays from dark matter annihilations which is most often studied is the central region of our galaxy. Recently, observations by the Atmospheric Cerenkov Telescopes (ACTs) HESS \cite{hess}, Whipple \cite{whipple} and Cangaroo \cite{cangaroo} have revealed the presence of a very bright gamma-ray source from this direction. The spectrum of this source has been measured in steadily increasing detail by the HESS collaboration \cite{hess}. Although the first HESS data from this source was not inconsistent with a spectrum from annihilating dark matter \cite{hessdark}, it is now becoming difficult to reconcile the HESS data with such a spectrum. Instead, it appears more likely that an astrophysical accelerator is responsible for this bright gamma-ray emission. As a result, future dark matter searches in this region will face a background that will be very challenging to overcome \cite{gabi}. 

Given this newly discovered background, it is important to consider other possible regions in which an observable rate of dark matter annihilation radiation may be generated. Such gamma-rays may appear as point sources external to our own galaxy, such as Andromeda (M31), M87 or the Large Magellanic Cloud~\cite{fornengo}, or as a diffuse spectrum generated by a large number of distant sources \cite{diffuse,closer}. Observable quantities of gamma-rays may also be generated in dark substructure within our own galactic halo. Again, this may appear as a diffuse spectrum from a large number of dark matter clumps \cite{mwclumps}, or may be dominated by a few of the most massive dwarf galaxies within the Milky Way, such as Draco, Sagittarius and Canis Major \cite{sarkar,tyler}.

In this article, we will discuss the prospects for detecting gamma-rays from dark matter annihilations in the dwarf galaxy Draco. We focus on this particular object for several reasons. First, of the most nearby and massive dwarf galaxies, the halo profile of Draco is the most tightly constrained by observations. Although other dwarfs may actually be brighter sources of dark matter annihilation radiation (this is likely for both Sagittarius and Canis Major \cite{sarkar}), the rates from these objects cannot be estimated with as much confidence. Second, since dwarf galaxies are dark matter dominated, containing very few baryons, gamma-ray searches for dark matter in these regions are very unlikely to be complicated by the presence of astrophysical sources. In light of the challenges faced for dark matter searches in the galactic center, this is clearly an important consideration.

A third reason that we chose to focus on Draco is the potentially exciting results of the CACTUS gamma-ray experiment. In recent conferences \cite{cactus}, the CACTUS collaboration has stated that they have detected an excess of $\sim$100 GeV gamma-rays from the direction of Draco. Although still preliminary, this result, if confirmed, would have dramatic implications for dark matter.

The remainder of this article is organized as follows. In the following section, we calculate the annihilation rate of dark matter in Draco, and the resulting gamma-ray flux. We then discuss the prospects for MAGIC and GLAST to detect this flux, and then lastly turn our attention to the possible detection of Draco by CACTUS, and the implications of such an observation for dark matter.

\section{Gamma-Rays From Dark Matter Annihilations In Draco}
%\label{basic}

Gamma-rays can be generated in dark matter annihilations through several processes. Most distinctive are those which result in mono-energetic spectral lines, $\chi \chi \rightarrow \gamma \gamma$, $\chi \chi \rightarrow \gamma Z$ or $\chi \chi \rightarrow \gamma h$. In most models, these processes only take place through loop diagrams, and thus the cross sections for such final states are quite suppressed, and lines are experimentally challenging to observe. 

A continuous spectrum of gamma-rays can also be produced through the fragmentation and cascades of most other annihilation products. The spectrum which results depends on the dominant annihilation modes. Parameterizations of the gamma-ray spectrum from dark matter annihilation can be found for several cases in Refs.~\cite{fornengo,buckley,closer}. 

The normalization of this spectrum depends on the dark matter profile of Draco.Assuming that the halo profile is approximately spherically symmetric, the
total annihilation rate in Draco within the radius $r_a$ is given by 
\beq
\Phi_A= \int_{r_{min}}^{r_a} dr \, 4\pi r^2 {\langle\sigma_Av\rangle\over 2}
\left({\rho(r)\over m_\chi}\right)^2,
\eeq
where $\langle\sigma_Av\rangle$ is the WIMP's annihilation cross section and $m_{\chi}$ is its mass. $\rho(r)$ is the density of dark matter at a radius $r$ from Draco's center. This annihilation rate leads to an isotropic flux of gamma-rays from Draco that is given by
\beq
F_\gamma={\Phi_A N_{\gamma} \over 4\pi D^2},
\eeq
where $D$ is the distance to Draco and $N_{\gamma}$ is the number of gamma-rays produced per annihilation in the energy range of a given detector. The distance to Draco has been determined to be $75.8\pm 0.7\pm 5.4$
kpc from an analysis of RR Lyrae variable stars \cite{cepheid}.

%%%

In contrast to the halo profile of the Milky Way galaxy, the properties of Draco's profile are somewhat constrained. For instance, it has been shown that Draco's halo has not been tidally stripped due to the interaction with the Milky Way halo~\cite{couchman}. Despite these constraints, however, current observations have been unable to determine the slope of Draco's inner halo profile, being equally consistent with a "cusped" halo profile, i.e. a density profile $\sim \rho^{\gamma}$, with $\gamma$ between $-0.5$ and $-1.5$ as given by N-body
simulations, or with a flat core ($\gamma\sim 0$). We will consider both of these possibilities.

%Another
%factor of uncertainty concerns the overall shape of the halo, where
%for example an elongation along the line of sight may boost the 
%expected annihilation signal. Also, the survival probability of the smallest
%halos, which are predicted to form in large numbers for WIMPS \cite{clumps}
%may be an important factor. Finally, the presence of an intermediate-size 
%black hole could influence the dark matter distribution by giving a spike
%of unknown density \cite{spike}. 

%Recently, there have been attempts to make specific models of the
%Draco dark matter halo and the star formation pattern using N-body
%simulations as a guide~\cite{couchman}. In that work it was shown that 
%an NFW-like profile could be fitted to the observations, but also a
%less steep profile (in fact, the authors of \cite{couchman} argue for
%$\alpha\sim 0.5$ as perhaps the best fit within large errors). We will explore each of these possibilities here.

Considering an NFW profile 
\beq
\rho(r)={\rho_0\over y\left(1+y\right)^2},
\eeq 
where $y=r/r_s$ is a dimensionless variable (and $r_s$ is the scaling 
radius), we arrive at
\beq
F_{\gamma}={\rho^2_0 r^3_s N_{\gamma} \over 3 m^2_{\chi} D^2}{\langle\sigma_Av\rangle\over 2}\left[{1\over \left(1+y_{min}\right)^3}-{1\over \left(1+y_a\right)^3}\right]
\label{nfwrate}
\eeq
where $y_{min}=r_{min}/r_s$ and $y_a=r_a/r_s$.

Alternatively, we can consider a halo with a flat central core, 
\beq
\rho(r)={\rho_0\over \left(1+y\right) \left(1+y^2\right)},
\eeq
which leads to
\beq
F_{\gamma}={\rho^2_0 r^3_s N_{\gamma} \over 4 m^2_{\chi} D^2}{\langle\sigma_Av\rangle\over 2}\left[{2+y_{min}+y^2_{min}\over 1+y_{min}+y^2_{min}+y^3_{min}}+\arctan(y_{min})-{2+y_a+y^2_a\over 1+y_a+y^2_a+y^3_a}-\arctan(y_a)\right].
\label{corerate}
\eeq
In table~\ref{rate}, we give numerical values for the NFW and cored profile cases, for some specific values of $y_{a}$ and $y_{min}$. These values allow us to rewrite Eqs.~\ref{nfwrate} and~\ref{corerate} as
\beq
F_{\gamma}={\rho^2_0 r^3_s N_{\gamma} \over 3 m^2_{\chi} D^2}{\langle\sigma_Av\rangle\over 2}\times A,
\label{rate}
\eeq
where $A$ is the value for a given profile found in table~\ref{ratetable}. There are several things to note about these results. Firstly, the annihilation rate does not depend critically on the outer radius integrated out to. Varying the outer radius between the scale radius, $r_a$, and much larger values, the overall annihilation rate varies only by about a factor of 2 for a cored profile, and much less for an NFW profile. Secondly, unlike in the case of the galactic center halo profile, the annihilation rate is not very much lower for a cored profile than for the NFW case. In particular, the rate is reduced only by a factor of 3 to 5, is the same values of $\rho_0$ and $r_s$ are adopted. 
%
%%%%%%%%%%%%%%
\begin{table}[t] 
\begin{center}
\begin{tabular}{|c|c|c|}  
\hline 
\raisebox{0mm}[4mm][2mm]  
Profile Type & $A (r_a=r_s)$ & $A (r_a \gg r_s)$   \\ 
\hline \hline
NFW  & 0.875 & 1.0 \\ 
\hline 
Core & 0.160 & 0.323  \\
\hline 
Cusp, $\gamma=1.1$ & 1.29 & 1.52  \\ 
\hline 
Cusp, $\gamma=1.2$ & 2.16 & 2.63  \\ 
\hline 
Cusp, $\gamma=1.3$ & 4.03 & 4.12  \\ 
\hline 
Cusp, $\gamma=1.4$ & 11.1 & 12.5  \\ 
\hline 
Cusp, $\gamma=1.45$ & 25.7 & 27.4  \\ 
\hline \hline
\end{tabular} 
\caption{Values of the parameter $A$, as used in Eq.\ref{rate}, for various halo halo profiles. In each case, $r_{min}=0$ was used. See text for more details.}
\label{ratetable}
\vspace{2mm}
\small
\end{center}
\end{table}
%%%%%%%%%%%
%
In addition to cored and NFW profiles, we have also shown in table~\ref{ratetable} results for profiles with a denser cusp. In the case that the density in the inner halo scales with $1/r^{\gamma}$ rather than $1/r$ as in the NFW case, we find somewhat larger annihilation rates, as expected. 

Up to these modest halo model-dependent variations in $A$, the annihilation rate and gamma-ray flux depends only on the quantity $\rho^2_0 r^3_s$, as well as the annihilation cross section and mass of the WIMP. Several observational constraints can be applied to Draco to constrain the quantities $\rho_0$ and $r_s$. Most important for our purposes are the constraints on the circular velocity, and the requirement that dark matter halos be abundant enough to account for the $\sim$20 dwarf spheroidal galaxies in the local group. Also important are the constraints arrived at by the requirement that the first stars formed in Draco at least 10 billion years ago, and that Draco's virial radius extends at least to the most distant stars ($\sim$1.2 kpc from Draco's center). Collectively, these constraints limit $r_s$ to values between 7 and 0.2 kpc, and $\rho_0$ to values between $10^7$ and $10^9 \,M_{\odot}/\rm{kpc}^3$ for the case of an NFW profile (the allowed ranges are somewhat smaller for the case of a cored profile) \cite{couchman}. Interestingly, the allowed regions in the $r_s$--$\rho_0$ plane are rather narrow strips, with a slope of roughly $\rho_0 \propto r^{-3/2}_s$  \cite{couchman}. This leads to little change in the quantity $\rho^2_0 r^3_s$ with variation of $\rho_0$ and $r_s$. For either an NFW or a cored halo profile, the allowed range of the quantity $\rho^2_0 r^3_s$ varies by only a factor of approximately 200 (0.03--6.3 $\times 10^{16} \, M^2_{\odot}/\rm{kpc}^3$)~\cite{couchman}.

Applying these constraints on $\rho_0$ and $r_s$ as well as the allowed range of $D$, we arrive at the following maximal and minimal gamma-ray fluxes from Draco:
\begin{eqnarray}
F^{\rm{max}}_{\gamma, \rm{NFW}} \approx 2.4 \times 10^{-10} \left({100\ {\rm GeV}\over m_\chi}\right)^2
\left({\langle \sigma_Av\rangle\over 3\cdot 10^{-26}\ {\rm cm}^3{\rm s}^{-1}}\right)\   \left({N_{\gamma}\over 10}\right)  {\rm cm}^{-2}{\rm s}^{-1}, \label{fluxes1}\\
F^{\rm{min}}_{\gamma, \rm{NFW}} \approx 9.8 \times 10^{-13} \left({100\ {\rm GeV}\over m_\chi}\right)^2
\left({\langle \sigma_Av\rangle\over 3\cdot 10^{-26}\ {\rm cm}^3{\rm s}^{-1}}\right)\   \left({N_{\gamma}\over 10}\right)  {\rm cm}^{-2}{\rm s}^{-1}, \label{fluxes2}\\
F^{\rm{max}}_{\gamma, \rm{core}} \approx 4.2 \times 10^{-11} \left({100\ {\rm GeV}\over m_\chi}\right)^2
\left({\langle \sigma_Av\rangle\over 3\cdot 10^{-26}\ {\rm cm}^3{\rm s}^{-1}}\right)\   \left({N_{\gamma}\over 10}\right)  {\rm cm}^{-2}{\rm s}^{-1}, \label{fluxes3}\\
F^{\rm{min}}_{\gamma, \rm{core}} \approx 3.5 \times 10^{-13} \left({100\ {\rm GeV}\over m_\chi}\right)^2
\left({\langle \sigma_Av\rangle\over 3\cdot 10^{-26}\ {\rm cm}^3{\rm s}^{-1}}\right)\   \left({N_{\gamma}\over 10}\right)  {\rm cm}^{-2}{\rm s}^{-1}, 
\label{fluxes4}
\end{eqnarray}
where $N_{\gamma}$ is the number of gamma-ray produced per annihilation in the energy range of the detector. In each case an angular radius around Draco of $1^{\circ}$ was considered (which impacts the value of $A$). 

This range of fluxes conservatively encompasses the range which can be
expected from annihilating dark matter in Draco given the current
constraints on its mass distribution. If less conservative profiles
are considered, such as denser cusps, fluxes larger by $1$--$2$ orders
of magnitude are possible. A smaller annihilation rate and gamma-ray
flux than the minimal cored profile case cannot be easily accommodated,
however. \footnote{Comparing our results to those of Evans, Ferrer and Evans \protect\cite{sarkar}, their fluxes fall near the center of our (maximal to minimal) allowed range (in log units).}

\section{Prospects for the MAGIC and GLAST Experiments}
%\label{magicglast}

The experimental technology employed by the field of gamma-ray astronomy is currently developing very rapidly. This holds true for both ground-based Atmospheric Cerenkov Telescopes (ACTs) and satellite-based gamma-ray detectors.

Of the currently operating ACTs, MAGIC is the best suited for observations of Draco. MAGIC's northern hemisphere location at La Palma in the Canary Islands allows it to observe in the direction of Draco (unlike HESS, for example). VERITAS, once operational, will also benefit from its northern hemisphere location. MAGIC is also designed to have a lower energy threshold ($\sim$50 GeV for overhead sources) than other ACTs ($\sim$200 GeV for HESS or VERITAS), which is very important for dark matter searches.  

\begin{figure}
\begin{center}

\resizebox{8cm}{!}{\includegraphics{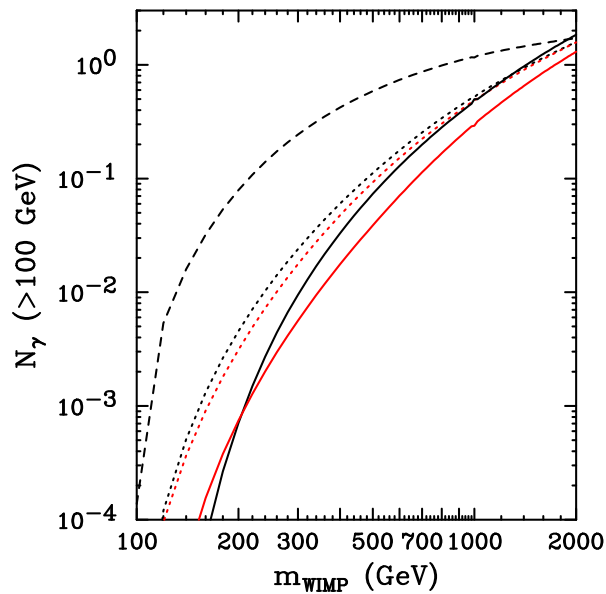}}
\caption{The number of gamma-rays produced per dark matter annihilation for several annihilation modes. Results for annihilations to $\tau^+ \tau^-$ are shown as a dashed line, $W^+W^-$ and $ZZ$ as dark and light (red) dotted lines, and $b \bar{b}$ and $t \bar{t}$ quarks as dark and light (red) solid lines.}
\label{ngamma100}
\end{center}
\end{figure}

From the latitude of the MAGIC telescope (22$^{\circ}$ north), Draco reaches zenith angles as small as 29$^{\circ}$. At this angle, an energy threshold of $\sim$100 GeV should be possible. Large numbers of gamma-rays are produced above this energy only for WIMPs considerably more massive. In figure~\ref{ngamma100}, the number of gamma-rays above 100 GeV per dark matter annihilation is shown as a function of the WIMP's mass for several dominant annihilation modes. This clearly shows the difficulty in observing a WIMP not much more massive than the energy threshold of MAGIC, or other ACT.

The primary background for ACTs is generated by hadronic cosmic rays. Fortunately, most of these showers can be identified and removed from the signal. In the energy range we are interested in, this background is roughly given by
\begin{equation}
\frac{dN_{\rm{bg}}}{dE_{\rm{bg}}} \approx \epsilon \times 10^{-5} \, \rm{GeV}^{-1} \,\rm{cm}^{-2} \, \rm{s}^{-1}\,\rm{sr}^{-1} \times \bigg(\frac{100\, \rm{GeV}}{E_{\rm{bg}}}\bigg)^{2.7},
\end{equation}
where $\epsilon$ is the fraction of hadronic showers which are misidentified as electromagnetic, which is on the order of 1\% for MAGIC. Integrating this above the 100 GeV threshold of MAGIC, and considering an effective area of $\sim 5 \times 10^8$ cm$^{2}$, this background accumulates at a rate of $\sim 100 \times \epsilon$ per hour over a $10^{-5}$ sr solid angle (approximately a $0.1^{\circ}$ by $0.1^{\circ}$ circle). 

\begin{figure}
\begin{center}

\resizebox{8cm}{!}{\includegraphics{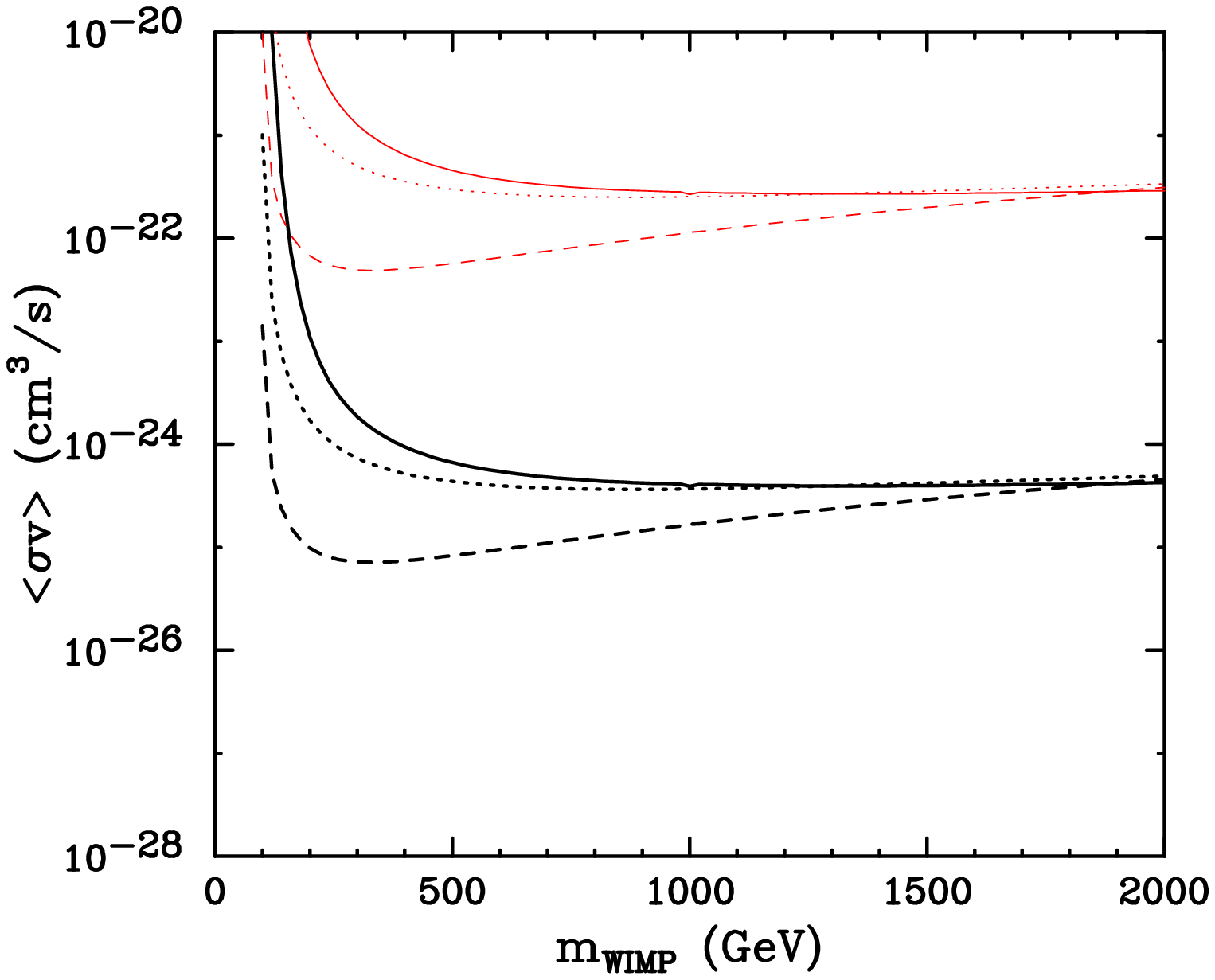}}
\resizebox{8cm}{!}{\includegraphics{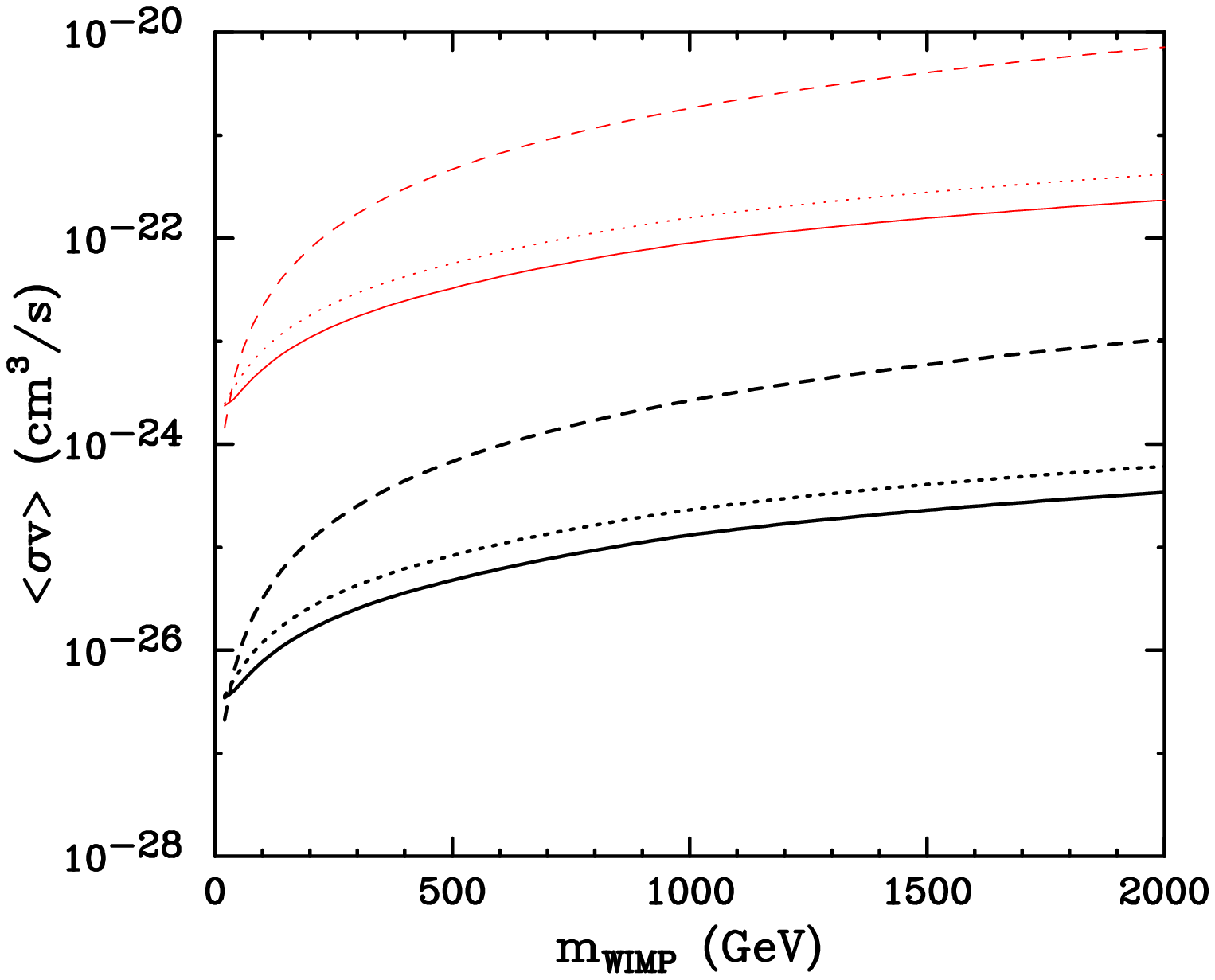}}
\caption{The sensitivity of MAGIC (left) and GLAST (right) to dark matter annihilation radiation from Draco. The lower, darker set of lines adopts our maximal NFW profile, while the upper, and lighter, set of lines adopts our most conservative, minimal cored profile (see Eqs.~\ref{fluxes1}-\ref{fluxes4}). For MAGIC, we have considered a $5\times 10^8$ cm$^2$ effective area, a solid angle of $10^{-5}$ sr, 40 hours of observation time, and 99\% hadronic seperation ($\epsilon=0.01$). For GLAST, we have considered a $10^4$ cm$^2$ effective area, a solid angle of $9 \times 10^{-5}$ sr and one year of observation time. In each frame, the solid, dotted and dashed lines correspond to annihilations to $b\bar{b}$, $W^+ W^-$ and $\tau^+ \tau^-$, respectively. All contours represent the cross section and mass required to generate a detection at the 5$\sigma$ level.
}
\label{reach}
\end{center}
\end{figure}

In the left frame of figure~\ref{reach} we show the sensitivity of MAGIC to dark annihilations in Draco for three representative annihilation modes, and two halo profiles.  We find that, for the case of a maximal NFW profile as discussed in the previous section, MAGIC will observe Draco with 5$\sigma$ significance only for dark matter particles with annihilation cross sections of $\sim 10^{-25}$ cm$^{3}$/s or higher. This is somewhat larger than the maximum value for a thermal relic with a density equal to the measured cold dark matter density. Relics which are generated non-thermally may have larger annihilation cross sections, however.

For the case of a satellite based experiment, such as GLAST, the prospects are quite different. GLAST is sensitive to gamma-rays down to 100 MeV, although to reduce the background, we will impose a threshold of 2 GeV, which is still far below that of MAGIC or other ACTs.  With no background from misidentified hadronic cosmic rays, the diffuse gamma-ray background is all that needs to be overcome by GLAST. In the direction of Draco, this background has been measured by EGRET to be approximately $3.3 \times 10^{-7}$ cm$^{-2}$ s$^{-1}$ sr$^{-1}$ for gamma-rays above 2 GeV. Over one year of observation, a solid angle of $0.3^{\circ}\times 0.3^{\circ}$, and a square meter effective area, this yields approximately 9 background events -- a rate considerably lower that for an ACT.  

In the right frame of figure~\ref{reach}, we show the reach of GLAST to dark matter annihilations in Draco. GLAST clearly does much better than MAGIC for dark matter annihilating to heavy quarks or gauge bosons. In particular, thermally generated WIMPs with an annihilation cross section of $\sim 3 \times 10^{-26}$ cm$^3$/s and lighter than $\sim$500--700 GeV are detectable by GLAST in the case of a maximal NFW halo profile. The prospects are less promising for GLAST in the case of annihilations to taus, however, as few low energy gamma-rays are produced in tau decays.

Some scenarios in which WIMPs are generated non-thermally can produce very large fluxes of gamma-rays from an object such as Draco. For example, in Anomaly Mediated Supersymmetry Breaking (AMSB) scenarios, the lightest supersymmetric particle is a nearly pure Wino which annihilates very efficiently to gauge boson pairs \cite{amsb}. If produced thermally, such a particle would only constitute a small fraction of the dark matter density. Non-thermal mechanisms can generate the observed relic density of Winos in AMSB scenarios, however \cite{nonthermal}.

\begin{figure}
\begin{center}

\resizebox{8cm}{!}{\includegraphics{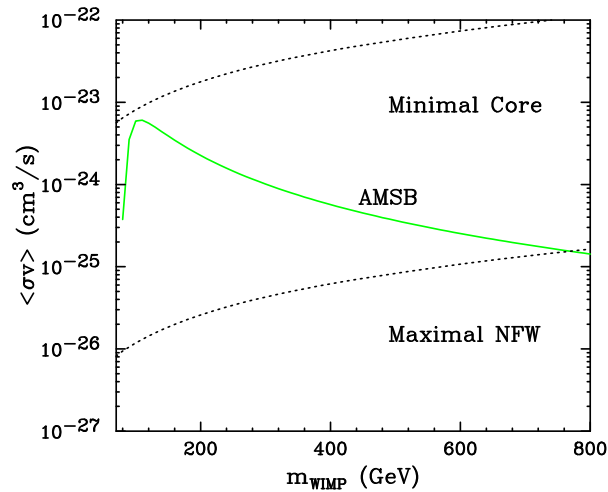}}
\caption{The sensitivity of GLAST to neutralino dark matter in Anomaly Mediated Supersymmetry Breaking (AMSB) scenarios. The upper and lower dotted lines show the sensitivity of GLAST to dark matter annihilations for GLAST (for a $5\sigma$ detection), adopting our minimal core and maximal NFW profiles, respectively (see Eqs.~\ref{fluxes1}-\ref{fluxes4}). The solid line represents the cross section predicted for a nearly pure-wino dark matter particle, as is present in AMSB scenarios. Even with a halo profile which produces the minimum possible annihilation rate in Draco (the minimal core), winos with a mass of 80 to 200 GeV could be potentially detected by GLAST. With other halo profiles, much heavier dark matter particles can be discovered.}
\label{amsb}
\end{center}
\end{figure}

In figure~\ref{amsb}, we show the sensitivity of GLAST to neutralino dark matter in AMSB scenarios. Even in the case of the halo profile which produces the {\it smallest} allowed annihilation rate in Draco (the minimal core profile) GLAST will be very close to being capable of detecting gamma-rays from neutralinos in the $\sim$80-200 GeV mass range. The absence of any signal from Draco seen by GLAST would thus exclude AMSB scenarios in this rather interesting mass range. If other halo profiles are considered, AMSB scenarios would likely be observable over a wide range of neutralino masses.

In addition to the continuum signals discussed so far, dark matter particles can also directly generated gamma-rays through loop-diagrams, leading to mono-energetic spectral lines. If such a line could be detected, it would represent a "smoking gun" for dark matter annihilation. Such lines are very difficult to detect, however. For neutralinos, for example, the cross section for gamma-ray lines is no larger than $\approx 10^{-28}$ cm$^{3}$/s, and can be up to 4 to 5 orders of magnitude smaller \cite{buckley,lines}. Also, only 1 or 2 gamma-rays are produced in each of these processes. With these considerations in mind, we can consider an optimal case: $m_{\chi}=30$ GeV, $\langle\sigma_Av\rangle = 10^{-28}$ cm$^{3}$/s, $N_{\gamma}=2$, and the maximal NFW profile. Together, this yields a flux of $\sim 2 \times 10^{-12}$ cm$^{-2}$ s$^{-1}$, or approximately half an event per year of exposure for GLAST. While even one event at $\sim$30 GeV would be intriguing from the direction of Draco, it would certainly not constitute a smoking gun of any kind. If a more steeply cusped halo profile were present, perhaps several line events could be seen by GLAST, but only in the most optimistic particle physics scenarios.

\section{The Observation of the CACTUS Experiment}
%\label{cactus}

CACTUS is a ground based gamma-ray telescope located near Barstow, California. It is largely sensitive to gamma-rays above $\sim$50 GeV, and has an effective area of up to $\sim$50,000 m$^2$ for $\sim$TeV gamma-rays. CACTUS employs an array of mirrors which were designed for solar observations rather than for gamma-ray astronomy, and therefore is not as optimally suited for gamma-ray detection as other ACTs. It is hoped, however, that its enormous effective area will make up for these disadvantages.

Very recently, the collaboration of the CACTUS experiment has, at a number of conferences, stated that they have detected an excess of gamma-rays from the direction of Draco \cite{cactus}. Although this result is still of a preliminary nature, it is interesting to consider the implications of such a detection if it is confirmed.

In 7 hours of observation, CACTUS detected approximately 30,000 events above background from Draco, of which roughly 7000 and 4000 were above 100 and 125 GeV, respectively~\cite{cactus}. To compare this with the predicted spectrum from annihilating dark matter, we convolve the injected spectrum with the energy dependent effective area of CACTUS, which has been parameterized as~\cite{cactus}
\begin{equation}
A_{\rm{eff}} \approx 47,000 \, \rm{m}^2 \, [1-e^{-0.014 \,(E_{\gamma}-39.6\, \rm{GeV})} +0.00025 \, E_{\gamma}(GeV)].
\end{equation}
%
%%%%%%%%%%%%%%
\begin{table}[t] 
\begin{center}
\begin{tabular}{|c|c|c|c|c|}  
\hline 
\raisebox{0mm}[4mm][2mm]   & Total &
$>$ 100 GeV & $>$ 125 GeV & EGRET \\ 
\hline \hline 
CACTUS Observation & 30,000 & 7000 & 4000 & --  \\ 
\hline \hline
600 GeV, $b\bar{b}$ & 30,000 & 9000 & 5000 & 290  \\ 
\hline 
500 GeV, $b\bar{b}$ & 30,000 & 7700 & 3900 & 400  \\ 
\hline 
400 GeV, $b\bar{b}$ & 30,000 & 6000 & 2700 & 630  \\ 
\hline \hline
400 GeV, $W^+W^-$ & 30,000 & 9200 & 5100 & 280  \\ 
\hline 
300 GeV, $W^+W^-$ & 30,000 & 7100 & 3500 & 470  \\ 
\hline 
200 GeV, $W^+W^-$ & 30,000 & 4000 & 1300 & 1100  \\ 
\hline \hline
300 GeV, $\tau^+\tau^-$ & 30,000 & 15,000 & 9500 & 2.8  \\ 
\hline 
200 GeV, $\tau^+\tau^-$ & 30,000 & 9200 & 4200 & 7.2  \\ 
\hline
150 GeV, $\tau^+\tau^-$ & 30,000 & 5000 & 1300 & 16  \\ 
\hline \hline
\end{tabular} 
\caption{The approximate energy distribution of events reported by CACTUS compared to the prediction from various annihilating dark matter scenarios. The CACTUS observations appear to be consistent with a $\sim$500 GeV dark matter particle annihilating to $b\bar{b}$, a $\sim$300 GeV dark matter particle annihilating to $W^+ W^-$, or a $\sim$200 GeV dark matter particle annihilating to $\tau^+ \tau^-$. In the last column, the number of events which EGRET should have seen is given for each case.}
\label{tab}
\vspace{2mm}
\small
\end{center}
\end{table}
%%%%%%%%%%%
Normalizing to the total rate seen by CACTUS, we can compare the energy distribution of events to that expected from annihilating dark matter. In table~\ref{tab} we show the number of events above 100 and 125 GeV expected for various dark matter masses and annihilation modes. The numbers reported by CACTUS appear to be consistent with the cases of a $\sim$500 GeV dark matter particle annihilating to $b\bar{b}$, a $\sim$300 GeV dark matter particle annihilating to $W^+ W^-$, or a $\sim$200 GeV dark matter particle annihilating to $\tau^+ \tau^-$. We emphasize, however, that systematic uncertainties in CACTUS's energy determination and understanding of backgrounds may modify these conclusions considerably. Despite these concerns, we conclude that annihilating dark matter with a mass in the 150-1000 GeV range appears to be consistent with the limited spectral information contained in the CACTUS signal from Draco.

Each of these scenarios require very high annihilation rates in Draco, however, which leads us to two potential problems. Firstly, a very cusped or spiked halo distribution would be needed to accommodate this rate -- roughly $10^3$ to $10^4$ times larger than the rate found for the maximal NFW model. To accommodate this, either a very large annihilation cross section (and a non-thermal production mechanism) or a very dense dark matter distribution (perhaps surrounding an intermediate mass black hole in Draco \cite{imbh}) would be required.

Secondly, the EGRET satellite observed this region, and has placed limits on 1--10 GeV gamma-rays from this region of the sky. For most of the scenarios shown in the table, this limit is violated.

Specifically, after imposing angular cuts of 1.71$^{\circ}$, 1.18$^{\circ}$ and 0.82$^{\circ}$ for gamma-rays between 1--2 GeV, 2--4 GeV and 4--10 GeV, respectively, EGRET actually observed 6 events, with an expected background of 4.1. Dark matter scenarios in the table which suggest that EGRET should have seen hundreds of gamma-rays from Draco are clearly inconsistent with this result. The exception to this problem, however, is for dark matter which annihilates to tau pairs. In this case, only a few events are expected to have been seen in EGRET. It is intriguing to note that this excess of $\sim$2 events observed by EGRET, although  certainly not statistically significant, is of the same order of magnitude as the rate expected for dark matter annihilating to $\tau^+ \tau^-$ given the spectrum reported by CACTUS. This annihilation mode may dominate, for example, in the case of a bino-like neutralino which annihilates through the exchange of a light stau.

If the CACTUS signal is in fact a gamma-ray spectrum (as opposed to a poorly understood background, for example), GLAST will detect thousands of gamma-rays from Draco, determining its spectrum in detail. MAGIC should also easily detect such a source with high significance. With MAGIC currently operating, we expect that it will not be long before the CACTUS signal is either confirmed or invalidated.

\section{Summary and Conclusions}
%\label{conclusions}

In this article, we have revisited the possibility of detecting gamma-rays produced in dark matter annihilations in the dwarf galaxy Draco. Draco is the most well constrained of the Milky Way's satellite galaxies, and therefore provides the best opportunity to make reliable predictions of dark matter annihilation rates and corresponding gamma-ray fluxes.

Using the constraints on the dark matter distribution of Draco put forth in Ref.~\cite{couchman}, we have calculated maximal and minimal annihilation rates (and corresponding gamma-ray fluxes), considering both a cusped (NFW) profile and a profile with flat core. The variation that we find in the annihilation rate between even these two extreme scenarios is less than three orders of magnitude. 

We then proceeded to compare these rates to the sensitivity of MAGIC and GLAST. MAGIC is a currently operating ground based gamma-ray telescope, while GLAST is a satellite based gamma-ray detector scheduled to be deployed in 2007. We find that while both MAGIC and GLAST have the ability to detect dark matter in Draco in some scenarios ({\it ie.} a maximal NFW profile, low WIMP mass and favorable annihilation modes), dark matter in Draco can go undetected by these experiments in other cases. In some extreme cases (such as the non-thermal generation of neutralinos in anomaly mediated supersymmetry breaking models, for example), however, the lack of a detection by GLAST of gamma-rays from Draco could successfully rule out models, even if the most conservative halo model were assumed.

Finally, we have also discussed the implications of the recent possible detection of Draco by the ground based gamma-ray detector, CACTUS. We find that to produce the signal reported by the CACTUS collaboration, an annihilation rate of dark matter in Draco is needed which is three to four orders of magnitude larger than can be accommodated for an NFW profile and an annihilation cross section consistent with thermally generated dark matter. Non-thermally produced dark matter and/or extremely high densities of dark matter in Draco would therefore be required to generate this signal. We also find that the CACTUS signal appears to be in conflict with the null results from the region by the EGRET experiment for most choices of the dark matter's dominant annihilation modes. If the dark matter almost entirely annihilates to tau pairs, however, this conflict can be (marginally) avoided. If CACTUS is in fact detecting this very large flux of gamma-rays from Draco, both MAGIC and GLAST should easily be able to confirm this result.

\bigskip

We would like to thank Gianfranco Bertone, Joakim Edsj\"o, Francis
Halzen, Peter Marleau, Karl Mannheim and Mani Tripathi for helpful
discussions. DH is supported by the US Department of Energy and by
NASA grant NAG5-10842. LB is supported by the Swedish Science
Research
Council (VR).

\end{document}